\documentclass{article} 
\usepackage{iclr2026_re-align_workshop,times}
\usepackage{booktabs}
\usepackage{tabularx}
\usepackage{geometry}
\usepackage{multirow}
\usepackage{wrapfig}
\usepackage{graphicx}
\usepackage{tcolorbox}
\usepackage{longtable}

\usepackage{amsmath,amsfonts,bm}

\def\eqref#1{equation~\ref{#1}}

\def\1{\bm{1}}

\DeclareMathAlphabet{\mathsfit}{\encodingdefault}{\sfdefault}{m}{sl}
\SetMathAlphabet{\mathsfit}{bold}{\encodingdefault}{\sfdefault}{bx}{n}

\usepackage{hyperref}
\usepackage{url}

\usepackage{xcolor}
\usepackage{soul}

\definecolor{hl-safety}{RGB}{255, 230, 230}
\definecolor{hl-reg}{RGB}{230, 240, 255}
\definecolor{hl-update}{RGB}{255, 250, 205}

\DeclareRobustCommand{\hlsafety}[1]{{\textbf{\sethlcolor{hl-safety}\hl{#1}}}}
\DeclareRobustCommand{\hlreg}[1]{{\textbf{\sethlcolor{hl-reg}\hl{#1}}}}
\DeclareRobustCommand{\hlupdate}[1]{{\textbf{\sethlcolor{hl-update}\hl{#1}}}}

\title{Aligning Sentence Embeddings \\ to Human Concepts via Sparse Autoencoders}

\author{Wonseok Shin, Songkuk Kim  \\
Yonsei University\\
\texttt{\{wonseok.shin,songkuk\}@yonsei.ac.kr} 
}

\iclrfinalcopy 

\begin{document}

\maketitle

\begin{abstract}
Dense sentence embeddings are fundamental to modern Retrieval-Augmented Generation (RAG) systems but suffer from a lack of interpretability due to feature superposition. This opacity hinders the alignment of retrieval processes with human intent, as the entangled representations are difficult to analyze or control. In this work, we propose a method to disentangle the dense representations of sentence transformers (e.g., E5) into human-interpretable concepts using Top-k Sparse Autoencoders (SAEs). We demonstrate that these disentangled features align with specific semantic, syntactic, and pragmatic categories. Furthermore, we introduce an activation steering mechanism that allows for precise intervention in the retrieval process. By clamping specific latent features, we show that it is possible to re-rank search results to better align with user constraints without retraining the backbone model. Our findings suggest that SAE-based decomposition offers a viable path toward transparent and steerable neural information retrieval.
\end{abstract}

\section{Introduction}

Neural sentence embeddings serve as the backbone of modern retrieval-centric applications, including Retrieval-Augmented Generation (RAG) pipelines \citep{lewis2020retrieval}, yet they suffer from limited interpretability \citep{lipton2018mythos}. According to the superposition hypothesis, dense vectors often compress sparse features into fewer dimensions, rendering them polysemantic \citep{arora2018linear, elhage2022toy}. This opacity creates a fundamental alignment gap \citep{luan2021sparse}, making it difficult to verify or intervene when retrieval processes diverge from human intent.

To bridge this gap, we propose applying Sparse Autoencoders (SAEs) \citep{ng2011sparse} to the final output of sentence encoders, such as the E5 model \citep{wang2022text}. While SAEs are typically used to study internal LLM residual streams \citep{bricken2023monosemanticity, cunningham2023sparse}, their potential to interpret dense retrievers remains underexplored. Specifically, we utilize a Top-k SAE architecture \citep{gao2024scaling} to map dense embeddings into a higher-dimensional, sparse latent space. This methodology allows us to decompose entangled representations into monosemantic features that are human-interpretable, expanding the 1,024-dimensional input into a significantly larger latent space to resolve feature superposition.

In this work, we demonstrate that Top-k SAEs can effectively align dense sentence embeddings with human conceptual frameworks. Our experiments on the WikiText-103-v1 dataset show that the learned latent features correspond to granular concepts. Furthermore, we go beyond static analysis to demonstrate semantic steering. By manually clamping specific latent neurons, we show that it is possible to control the retrieval mechanism—filtering out unwanted concepts without retraining the backbone model. This capability addresses the limitations of macro-steering by enabling micro-level control over semantic features. This work suggests that SAE-based decomposition is a viable path toward transparent, aligned, and steerable neural search systems.

\section{Methodology}

We propose a framework to disentangle and align the dense representations of neural retrievers using a Top-k Sparse Autoencoder (SAE). This approach maps entangled dense embeddings into a high-dimensional sparse latent space where individual dimensions represent monosemantic concepts.

\subsection{Top-k Sparse Autoencoder Architecture}

We adopt the Top-k SAE architecture to overcome the limitations of traditional $L_1$-regularized autoencoders. While $L_1$ penalties effectively induce sparsity, they often introduce a shrinkage bias that suppresses feature activation magnitudes, degrading reconstruction quality. \citep{bricken2023monosemanticity, bussmann2024batchtopk} The Top-k approach addresses this by directly enforcing a hard constraint \citep{makhzani2015winner}: for every input embedding $\mathbf{x}$, only the $k$ most active latent neurons are retained, while the rest are set to zero.

Our model projects the input embedding into a latent space $\mathbb{R}^{d_{latent}}$ (where $d_{latent} \gg d_{model}$), applies the Top-k activation function, and reconstructs the original vector from these sparse features. Following the approach of \citet{gao2024scaling}, we tie the encoder and decoder weights to stabilize training. This ensures that the model learns a consistent and interpretable decomposition of the semantic space without the trade-offs inherent in soft sparsity constraints.

For the detailed mathematical formulation, and the auxiliary training objective used to mitigate dead neurons, please refer to Appendix \ref{app:sae_details}.

\subsection{Latent Feature Steering via Clamping}

A key contribution of our work is the ability to intervene in the retrieval process via latent feature clamping. Since each dimension $i$ of $\mathbf{z}$ corresponds to an interpretable concept, we can surgically modify the latent vector to filter out unwanted semantic attributes from the retrieval results.

Given an input query $\mathbf{x}$, we compute its sparse representation $\mathbf{z}$. We define a clamping operation $C(\mathbf{z}, \mathcal{I}_{\text{clamp}})$ where $\mathcal{I}_{\text{clamp}}$ is the set of target neuron indices to deactivate:
\begin{equation}
    \mathbf{z}_{\text{steered}}^{(i)} = 
    \begin{cases} 
    0 & \text{if } i \in \mathcal{I}_{\text{clamp}} \\
    \mathbf{z}^{(i)} & \text{otherwise}
    \end{cases}
\end{equation}
By strictly zeroing out specific activations, this operation effectively removes corresponding concepts without affecting other semantic components. The steered embedding $\hat{\mathbf{x}}_{\text{steered}} = \mathbf{W}_{\text{dec}}\mathbf{z}_{\text{steered}}$ is then used for similarity search, enabling precise, inference-time control over the retrieval mechanism without fine-tuning the backbone model.

\subsection{Automated Feature Annotation Pipeline}

To scale the interpretation of thousands of latent features, we utilize an automated pipeline powered by GPT-4o-mini. The process follows three stages:

\textbf{1. Orthogonality-based Filtering.} We prioritize distinct features by calculating the Decoder Orthogonality (DO) for each neuron. Only neurons with a mean pairwise cosine similarity below a strict threshold  are selected, ensuring that the annotated features are geometrically and semantically well-separated.

\textbf{2. Context Extraction.} For each target neuron, we retrieve the top-$N$ sentences (e.g., $N=10$) from the corpus that yield the highest activation values, providing a representative context of the learned pattern.

\textbf{3. LLM Labeling with Coherence Check.} An LLM acts as a linguist to identify common patterns across the retrieved sentences. Crucially, a Coherence Check is enforced: if the sentences lack a clear common thread, the neuron is classified as \textit{Unlabelable}. Otherwise, the LLM generates a specific \textbf{Label} and assigns a \textbf{Category} (e.g., Entity, Topic, Syntactic, or Pragmatic). Full details of the annotation prompts are provided in Appendix \ref{app:prompt}.

\section{Experiments}

\subsection{Experimental Setup}

\textbf{Dataset \& Model.} We utilize the WikiText-103-v1 dataset \citep{merity2017pointer} as our primary corpus. To ensure the quality of semantic representations, we applied a rigorous preprocessing pipeline involving sentence segmentation and length-based filtering, resulting in a final dataset of approximately 3.8 million sentences. Further details on the data preprocessing are provided in Appendix \ref{app:preprocessing}. 

For the backbone model, we employ \texttt{intfloat/e5-large-v2}, which generates 1,024-dimensional dense embeddings.

\textbf{SAE Configuration.} We train a Top-k SAE with a latent expansion factor of $12\times$, resulting in a latent dimension of $d_{latent} = 12,288$ given the input dimension of $d_{model}=1,024$. The sparsity level is fixed at $k=32$. To mitigate the dead neuron problem and ensure efficient feature utilization, we employ an auxiliary loss with $\alpha=0.1$.

\subsection{Quantitative Analysis: Disentanglement Quality}

\begin{wrapfigure}{r}{0.45\textwidth} 
     \centering
    \vspace{-12pt} 
    \includegraphics[width=0.45\textwidth]{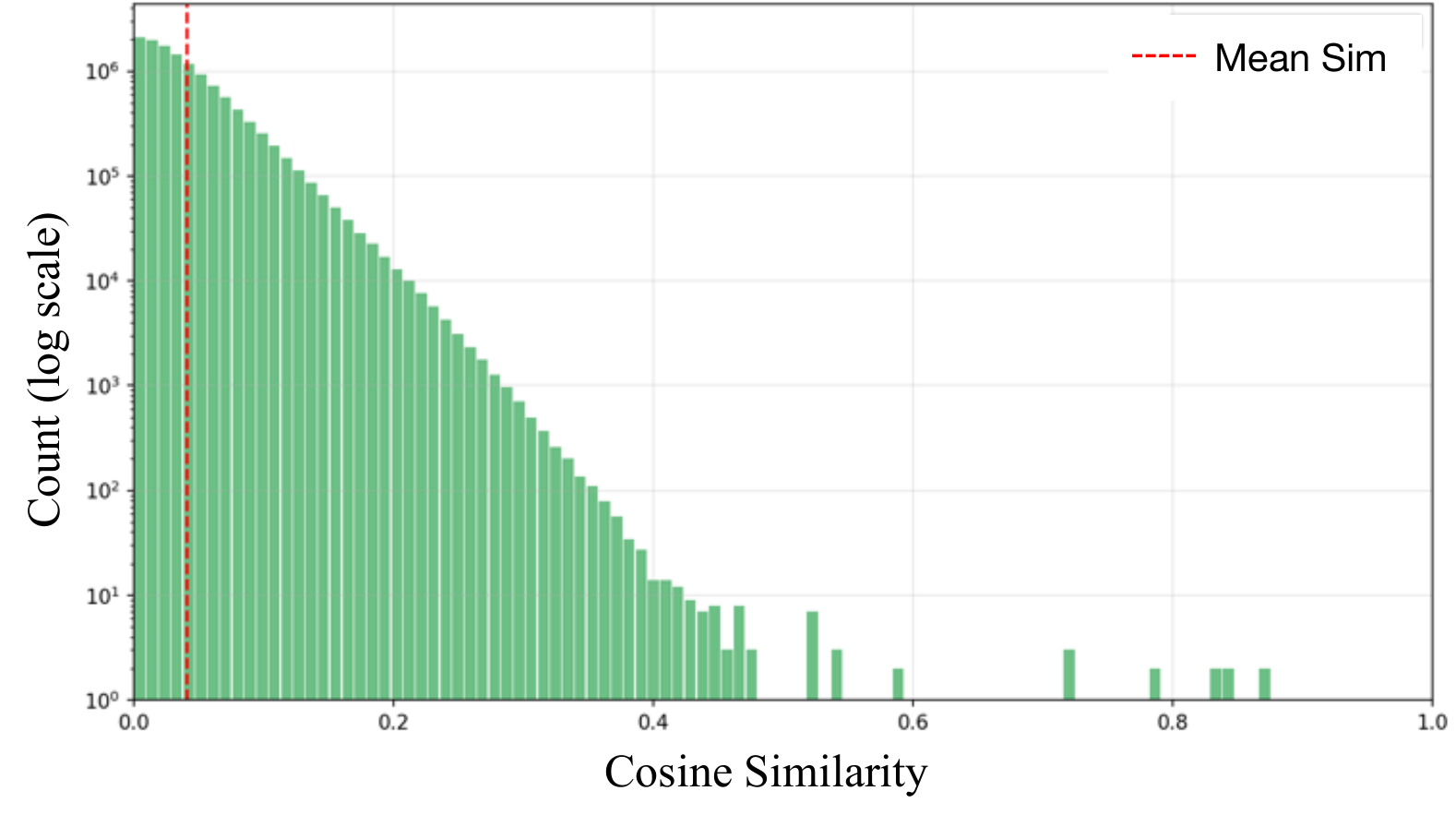} 
    \vspace{-18pt}
    \caption{Neuron activation frequency distribution (log-scale) on the Wikipedia corpus. }
    \label{fig:orthogonality_dist}
  \vspace{-7pt} 
\end{wrapfigure}

To verify that the SAE effectively disentangles the dense embedding space, we analyze two key metrics: \textbf{Decoder Orthogonality} and \textbf{Reconstruction Fidelity}. For detailed mathematical definitions of these metrics, please refer to Appendix \ref{app:metrics}. Based on the trade-off between semantic separability and information retention, we selected the model with a latent dimension of $d_{latent}=12,288$ ($12\times$ expansion) for our subsequent steering experiments.

\textbf{Decoder Orthogonality (DO).} A key prerequisite for interpretability is that learned features should represent distinct, non-overlapping concepts. We quantify this by measuring the mean pairwise cosine similarity between the columns of the decoder matrix $\mathbf{W}_{\text{dec}}$. As presented in Table \ref{tab:sae_performance}, our $12\times$ model achieves a remarkably low mean orthogonality score of 0.0408. Figure \ref{fig:orthogonality_dist} further illustrates the distribution of these pairwise similarities. The log-scale histogram reveals that the vast majority of feature pairs exhibit near-zero similarity, confirming that the learned semantic directions are structurally independent rather than redundant.

\begin{table}[h]
\centering
\begin{tabular}{lcccc}
\toprule
\textbf{Model} & 
\textbf{Explained Variance} & 
\textbf{Mean DO} & 
\textbf{Dead Neurons (\%)} \\
\midrule
$d_{latent}=4,096$  & 0.9136 & 0.0371 & 0.49 \\
$d_{latent}=8,192$  & 0.9211 & 0.0389 & 1.01 \\
$d_{latent}=12,288$ & 0.9259 & 0.0408 & 0.74 \\ 
\bottomrule
\end{tabular}%
\caption{Reconstruction performance and sparsity metrics of Top-k SAEs. The expansion factor denotes the ratio of the latent dimension to the model dimension ($d_{model}=1,024$). All models use $k=32$.}
\label{tab:sae_performance}
\end{table}

\begin{wrapfigure}{r}{0.5\textwidth} 
     \centering
    \vspace{0pt} 
    \includegraphics[width=0.5\textwidth]{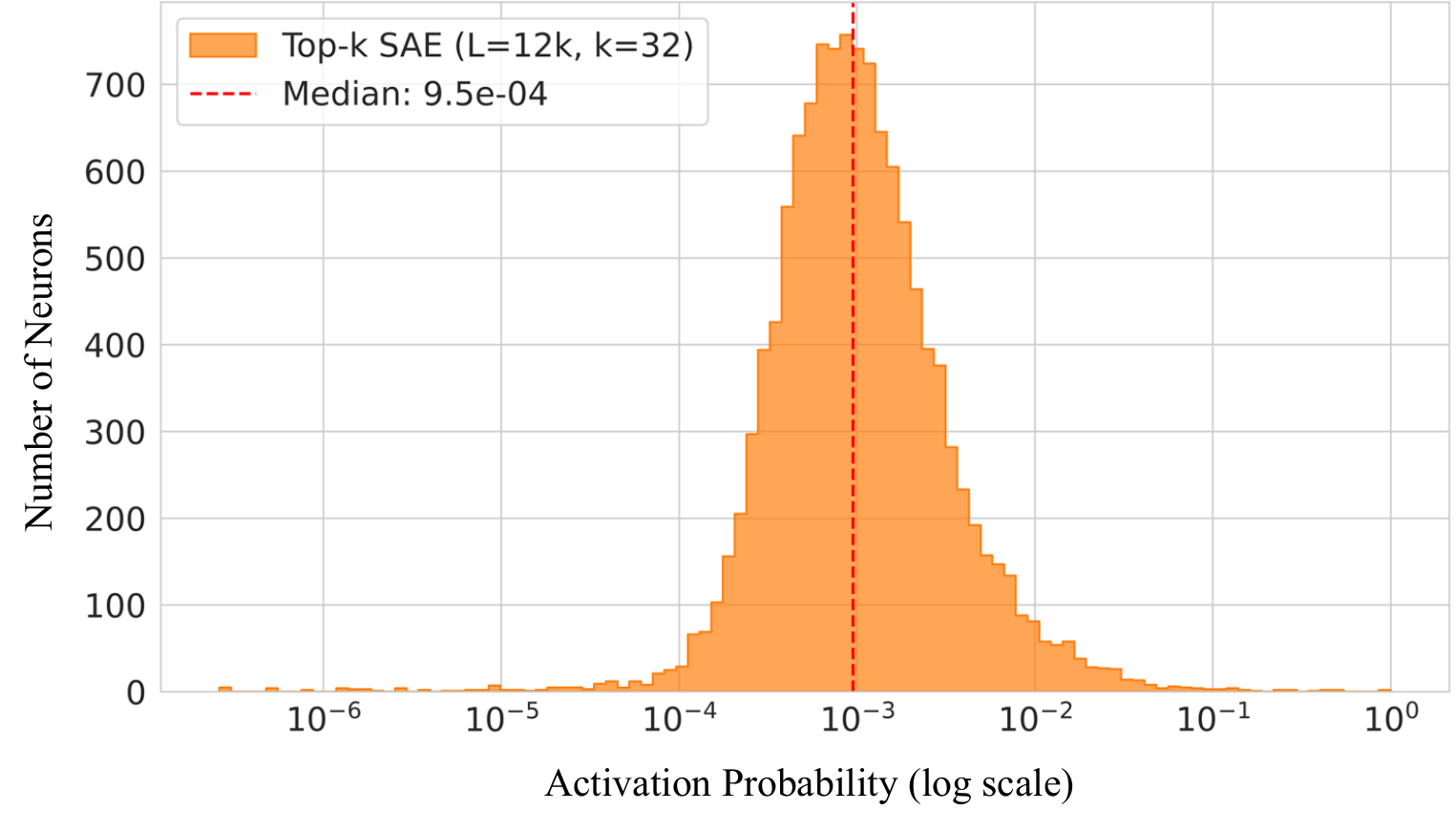} 
    \vspace{-18pt}
    \caption{Neuron activation frequency distribution on the Wikipedia corpus. The characteristic long-tail indicates that the Top-k SAE effectively utilizes its expanded latent space to maintain feature sparsity and vitality.}
    \label{fig:freq_dist}
  \vspace{-10pt} 
\end{wrapfigure}

\textbf{Reconstruction Fidelity.} Despite the aggressive sparsity constraint ($k=32$), the model maintains high fidelity with an Explained Variance (EV) of 0.9259. This corresponds to a Fraction of Variance Unexplained (FVU) of less than 0.08, demonstrating that the sparse decomposition effectively preserves the essential semantic information of the original dense embeddings.

\textbf{Feature Vitality and Sparsity.} We analyze the activation frequency across the corpus to evaluate latent space utilization. As shown in Figure \ref{fig:freq_dist}, the model exhibits a healthy long-tailed distribution with zero dead neurons (0.74\% mortality). The median activation frequency of $\approx 10^{-5}$ confirms that features capture granular, rare contexts rather than generic noise, providing a robust foundation for subsequent steering experiments.

\subsection{Analysis of Auto-Labeled Features}

We applied our automated pipeline to the trained Top-k SAE using GPT-4o-mini. Out of the neurons satisfying the orthogonality threshold, the LLM successfully generated coherent labels for the majority of features, confirming that the low-orthogonality criterion effectively filters for monosemantic concepts.

\textbf{Granularity of Concepts.} Unlike standard clustering which often stops at broad topics (e.g., ``Politics'' or ``Arts''), the auto-labeled features exhibit remarkable granularity. For instance, the model captures specific abstract topics such as ``Critical Reception of Media'' (\#7400) rather than general ``Reviews,'' and distinguishes specific syntactic tokens like ``Reference to Number 21'' (\#230) from general numeric values. As shown in Table \ref{tab:auto_labels}, these features span across diverse categories including specific entities, pragmatic functions, and syntactic patterns.

\begin{table}[h]
\centering
\small 
\renewcommand{\arraystretch}{1.2} 
\begin{tabularx}{\textwidth}{c l p{2.5cm} X}
\toprule
\textbf{ID} & \textbf{Category} & \textbf{Label} & \textbf{Representative Top Sentences} \\
\midrule

\#11 & Entity-Specific & \textit{Rosetta \& Rose Entities} & 
1) "Rosetta is the name of a lightweight dynamic translator..." \newline
2) "Okafor attended Rosemont Elementary." \\
\midrule

\#7400 & Topic-Specific & \textit{Critical Reception of Media} & 
1) "Despite the overwhelming negative reviews, the film did receive some positive feedback." \newline
2) "Overall, reception of the album was positive." \\
\midrule

\#487 & Pragmatic & \textit{Quotations and Citations} & 
1) "According to Mahatma Gandhi :" \newline
2) "The Liber Eliensis described the situation as follows :" \\
\midrule

\#230 & Syntactic/Token & \textit{Reference to Number 21} & 
1) "His uniform number was 21." \newline
2) "There were 21 fatalities." \\

\bottomrule
\end{tabularx}
\caption{Selected examples of auto-labeled features representing diverse categories. The model captures highly granular concepts, ranging from specific entities to syntactic tokens and pragmatic functions. Each feature is illustrated with top-activating sentences.}
\label{tab:auto_labels}
\end{table}

\subsection{Latent Steering for Alignment}

Building on the interpretability confirmed above, we demonstrate the capability to align retrieval results with human intent via activation steering. By manually clamping specific latent neurons, we can surgically intervene in the retrieval process to modify semantic priorities without retraining the backbone model.

Consider a complex query: \textit{``The \hlupdate{updated} \hlsafety{safety} \hlreg{guidelines} applied only to heavy machinery operators.''} We intervene by identifying and suppressing a specific semantic dimension:

\begin{itemize}
    \item \textbf{Original Retrieval:} The baseline retrieval prioritizes the entangled context of safety and labor.
    \begin{itemize}
        \item \textit{This was largely done in \hlsafety{consideration of safety grounds} and usually applied to those conducting maintenance or the repair of equipment.}
        \item \textit{To ensure \hlsafety{safety} during the Stack period, the organizers maintained a perimeter around the working area, and allowed only \hlsafety{safety-trained} students through.}
    \end{itemize}
    
    \item \textbf{Steering Intervention:} We identified neuron \#4940 (\textit{Industrial Safety}) via our auto-labeling pipeline and clamped its activation to zero ($z_{4940} = 0$).
    
    \item \textbf{Steered Retrieval:} Neutralizing the `safety' component shifts results toward the regulatory scope; the original top result dropped to 9th place, while regulatory sentences ascended:
    \begin{itemize}
        \item \textit{The final rule adopted several \hlupdate{changes} to the HOS \hlreg{regulations}, including a new \hlreg{provision} requiring drivers to take a rest break during the work day under certain circumstances.}
        \item \textit{The policy has been \hlupdate{changed} so \hlreg{permits} are only required for large scale film, video and photography requiring 10 person crews .}
    \end{itemize}
\end{itemize}

These results demonstrate that Top-k SAEs effectively bridge dense vectors and human concepts. By modulating these semantic atoms, we align neural retrieval with specific intent, offering a principled framework for steerable, transparent search. See Appendix \ref{app:more_cases} for further case studies.

\section{Conclusion and Future Work}

In this work, we presented a framework for aligning the latent representations of dense retrievers with human conceptual structures utilizing Top-k Sparse Autoencoders. We demonstrated that applying the Top-k architecture allows for the successful decomposition of 1,024-dimensional embeddings into granular, monosemantic features, effectively bridging the gap between high-dimensional vector spaces and human interpretability.

Our automated analysis pipeline confirmed that these features correspond to specific entities, topics, and pragmatic functions. Crucially, we demonstrated that this disentanglement enables latent steering, which serves as a mechanism to intervene in the retrieval process and re-rank results according to user intent without model retraining.

\textbf{Limitations.} Our study focuses on a single backbone model (E5-large) and English benchmarks. Furthermore, we acknowledge that our hyperparameter selection is not exhaustive. While prior research indicates that scaling the latent dimension—often accompanied by increasing $k$—can improve performance up to the point of model collapse, we did not optimize for the ideal latent dimension or sparsity level. Since our primary objective was an exploratory investigation to verify if sentence embeddings could be decomposed into interpretable concepts and effectively steered, we maintained a fixed sparsity of $k=32$. Consequently, the current performance may not represent the upper bound of the proposed framework.

\textbf{Future Work.} Our primary focus for future research is to integrate this steering mechanism into RAG  pipelines to dynamically filter biased or unsafe concepts at inference time. We also plan to extend this methodology to multilingual settings to investigate cross-lingual conceptual alignment.

\newpage

\bibliography{iclr2026_conference}
\bibliographystyle{iclr2026_conference}

\section*{Disclosure of AI Usage}
We acknowledge the use of Gemini-3 for linguistic polishing to improve the manuscript's readability. Furthermore, GPT-4o-mini was employed within our research pipeline to perform automated feature annotation (Section 3.3). The authors maintain full responsibility for the content and ensure that all AI-generated outputs were critically evaluated and verified by the authors.

\newpage
\appendix
\section{Qualitative Analysis of Learned Latent Features}
\label{app:qualitative_analysis}

In this section, we provide a detailed examination of the features learned by our Top-k Sparse Autoencoder. Table \ref{tab:sae_features} presents a curated list of neurons, their human-interpreted labels, and representative activating sentences from the WikiText-103-v1 dataset. These examples demonstrate the model's ability to disentangle diverse linguistic and semantic concepts, ranging from specific entities to abstract topics and structural patterns.

\begin{table}[h]
\centering
\small 
\renewcommand{\arraystretch}{1.2} 
\begin{tabularx}{\textwidth}{c l p{3.2cm} X}
\toprule
\textbf{ID} & \textbf{Label} & \textbf{Description} & \textbf{Representative Activating Sentences} \\
\midrule

\#9689 & Music Critical Reception & Summaries of professional reviews for songs or albums, typically citing "music critics". & 
1) "Upon its release, the track garnered \textbf{positive reviews from music critics}, who praised the song's composition..." \newline
2) "The song received \textbf{generally favorable reviews from music critics} who commended Beyoncé's vocal performance..." \newline
3) "The song received \textbf{generally mixed reviews from music critics}... both praising and criticizing the inclusion of the sample..." \\
\midrule

\#311 & Disaster Impact & Descriptions of severe damage or "hardest hit" locations from storms/events. & 
1) "The \textbf{hardest hit areas} were in Jackson and Victoria counties where the \textbf{heaviest rains} fell." \newline
2) "\textbf{Damage was heaviest} in the northern offshore islands and in the northern portion..." \newline
3) "Coastal areas were \textbf{hardest hit}." \\
\midrule

\#11440 & Entity: The Ramones & Specific references to the punk rock band 'The Ramones' and members. & 
1) "The \textbf{Ramones} were an American punk rock band that formed in... Forest Hills, Queens." \newline
2) "\textbf{Ramones} is the debut studio album by the American punk rock band the \textbf{Ramones}..." \newline
3) "All songs were written by the \textbf{Ramones}, except where noted." \\
\midrule

\#8870 & Rhyme Scheme & Discussions of poetic structure, internal rhymes, and nursery rhymes. & 
1) "The \textbf{rhyme scheme} is AABB, or AA, B, CC, CB, B, B when accounting for internal \textbf{rhyme}." \newline
2) "Most \textbf{rhyme schemes} are described using letters that correspond to sets of \textbf{rhymes}..." \newline
3) "The poem's \textbf{rhyme scheme} is rhyming couplets rendered aa bb cc dd ee aa." \\
\midrule

\#1427 & "Ajax" (Polysemantic) & Activates for AFC Ajax (Football), HMS Ajax (Ship), and Mythology. & 
1) "\textbf{Ajax} won the tie 4–1 on aggregate to progress to the second round." (Sport) \newline
2) "The ship was renamed \textbf{Ajax} on 15 June 1869." (Naval) \newline
3) "\textbf{Ajax} focuses on the proud hero of the Trojan War, Telamonian \textbf{Ajax}..." (Mythology) \\
\midrule

\#8408 & Phrase: "There is" & Rhetorical or existential statements starting with "There is/are". & 
1) "\textbf{There has to be} something better." \newline
2) "What else \textbf{is there}?" \newline
3) "\textbf{There is} currently no authoritative voice classification system..." \\

\bottomrule
\end{tabularx}
\caption{Selected interpretable features learned by the Top-k SAE ($d_{latent}=12288$). We present three representative activating sentences per neuron to demonstrate semantic consistency. }
\label{tab:sae_features}
\end{table}

\section{Detailed SAE Architecture and Training Objective}
\label{app:sae_details}

In this section, we provide the mathematical formulation of our Top-k Sparse Autoencoder and the training objectives employed.

\subsection{Forward Pass Formulation}
Let $\mathbf{x} \in \mathbb{R}^{d_{model}}$ be the dense sentence embedding generated by the backbone model. The encoding process consists of a linear transformation followed by a Top-k non-linearity:
\begin{equation}
    \mathbf{z}_{\text{pre}} = \mathbf{W}_{\text{enc}}(\mathbf{x} - \mathbf{b}_{\text{dec}})
\end{equation}
\begin{equation}
    \mathbf{z} = \text{TopK}(\text{ReLU}(\mathbf{z}_{\text{pre}}), k)
\end{equation}
where $\mathbf{W}_{\text{enc}} \in \mathbb{R}^{d_{latent} \times d_{model}}$ is the encoder weight matrix, and $\mathbf{b}_{\text{dec}}$ is the decoder bias. The $\text{TopK}(\cdot, k)$ operation keeps only the $k$ largest values of the pre-activations and zeroes out the rest, ensuring a fixed sparsity level.

The reconstruction $\hat{\mathbf{x}}$ is obtained via the decoder:
\begin{equation}
    \hat{\mathbf{x}} = \mathbf{W}_{\text{dec}}\mathbf{z} + \mathbf{b}_{\text{dec}}
\end{equation}
We enforce tied weights such that $\mathbf{W}_{\text{enc}} = \mathbf{W}_{\text{dec}}^T$, which acts as a regularization technique to improve training stability and feature coherence.

\subsection{Training Objective with AuxK Loss}
The primary objective is to minimize the reconstruction error, measured by the Mean Squared Error (MSE):
\begin{equation}
    \mathcal{L}_{\text{recon}} = ||\mathbf{x} - \hat{\mathbf{x}}||_2^2
\end{equation}

A common challenge in training large-scale sparse autoencoders is the "dead neuron" problem, where certain latent features never activate across the entire dataset. To mitigate this, we incorporate an auxiliary loss term, $\mathcal{L}_{\text{aux}}$, often referred to as the AuxK loss. This objective encourages inactive neurons to predict the residual error $(\mathbf{x} - \hat{\mathbf{x}})$ of the main autoencoder, effectively "reviving" them by forcing them to explain the information missed by the currently active features.

The total loss function is defined as:
\begin{equation}
    \mathcal{L}_{\text{total}} = \mathcal{L}_{\text{recon}} + \alpha \mathcal{L}_{\text{aux}}
\end{equation}
where $\alpha$ is a hyperparameter that controls the contribution of the auxiliary task. In our experiments, we set $\alpha$ to a non-zero value (e.g., 0.1) to proactively prevent feature collapse.

\section{Dataset Preprocessing Details}
\label{app:preprocessing}

We constructed our training corpus using the WikiText-103-v1 dataset, which consists of over 100 million tokens extracted from verified ``Good and Featured'' articles on English Wikipedia. To adapt this document-level dataset for sentence embedding tasks, we performed the following preprocessing steps:

\begin{enumerate}
    \item \textbf{Sentence Segmentation:} We split the raw articles into individual sentences using the NLTK sentence tokenizer \citep{kiss2006unsupervised, loper2002nltk} as a standard natural language processing tool.
    \item \textbf{Filtering:} To remove noise and ensure semantic completeness, we applied several filters:
    \begin{itemize}
        \item We removed sentences shorter than 5 words to exclude section headers, list items, and fragmented text.
        \item We excluded extremely long sentences (e.g., $>128$ words) to prevent excessive truncation by the backbone model's tokenizer.
        \item We filtered out non-natural language artifacts, such as URLs and wiki-markup boilerplate.
    \end{itemize}
\end{enumerate}

After this process, the final dataset consists of 3,837,611 unique sentences, which serve as the inputs for training our Top-k Sparse Autoencoders.

\section{Evaluation Metrics}
\label{app:metrics}

Following the framework of \citet{zaigrajew2025interpreting}, we employ quantitative metrics to assess the quality of learned sparse representations.

\subsection{Explained Variance (EV) and FVU}
To measure reconstruction quality, we report the Explained Variance (EV). This metric is directly related to the Fraction of Variance Unexplained (FVU), where $\text{EV} = 1 - \text{FVU}$.

\begin{equation}
    \text{EV} = 1 - \frac{\sum_{i=1}^{N} ||\mathbf{x}_i - \hat{\mathbf{x}}_i||_2^2}{\sum_{i=1}^{N} ||\mathbf{x}_i - \bar{\mathbf{x}}||_2^2}
\end{equation}

A value close to 1.0 (or FVU close to 0.0) indicates that the sparse decomposition retains nearly all information present in the original dense embeddings.

\subsection{Decoder Orthogonality (DO)}
To evaluate the disentanglement of the learned features, we measure the orthogonality of the decoder weights. In an ideal disentangled representation, distinct latent features should correspond to distinct (orthogonal) directions in the semantic space.

Let $\mathbf{w}_i$ denote the $i$-th column vector of the decoder matrix $\mathbf{W}_{\text{dec}} \in \mathbb{R}^{d_{model} \times d_{latent}}$, representing the direction of the $i$-th latent feature. We first normalize all column vectors to unit length: $\hat{\mathbf{w}}_i = \frac{\mathbf{w}_i}{||\mathbf{w}_i||_2}$. We then compute the mean absolute pairwise cosine similarity between all unique pairs of features:

\begin{equation}
    \mathcal{M}_{\text{ortho}} = \frac{1}{d_{latent}(d_{latent}-1)} \sum_{i \neq j} |\langle \hat{\mathbf{w}}_i, \hat{\mathbf{w}}_j \rangle|
\end{equation}

where $\langle \cdot, \cdot \rangle$ denotes the dot product. A lower $\mathcal{M}_{\text{ortho}}$ score indicates that the features are more independent and less redundant. A score of 0 implies perfect orthogonality among all feature directions.

\newpage

\section{Automated Labeling Prompt}
\label{app:prompt}

To ensure transparency and reproducibility, we provide the full system and user prompts used for the automated feature annotation pipeline powered by GPT-4o-mini. The prompt was designed to enforce strict coherence checks and ensure specific, granular labels.

\begin{figure}[h]
\centering
\begin{tcolorbox}[colback=gray!5, colframe=gray!40, title=\textbf{System Prompt}]
\small
You are an expert linguist analyzing sentences that activate a specific SAE neuron. Your goal is to find a \textbf{specific, coherent pattern}.

\textbf{CRITICAL INSTRUCTION: COHERENCE CHECK} \\
Before labeling, check if the sentences share a strong semantic, syntactic, or pragmatic connection.
\begin{itemize}
    \item If the sentences describe completely unrelated topics with no common thread, you MUST classify it as \textbf{`Unlabelable'}.
    \item Do NOT force a label if the pattern is weak or ambiguous.
\end{itemize}

If coherent, identify the \textbf{most specific} common thread possible (Entity, Issue, Pragmatic Role). All outputs must be in \textbf{English}.
\end{tcolorbox}
\vspace{0.2cm}
\begin{tcolorbox}[colback=gray!5, colframe=gray!40, title=\textbf{User Prompt}]
\small
Here are the sentences that most strongly activate Neuron \#\{neuron\_idx\}. The number in brackets is the activation score (higher is stronger).

\textbf{[Sentences]} \\
\{List of top-20 activating sentences with scores\}

Analyze the patterns deeply and output the result in JSON format with the following keys:

\begin{enumerate}
    \item \textbf{``label''}: 
    \begin{itemize}
        \item If coherent: A \textbf{highly specific} label in English (3-4 words). e.g., ``Opposition to Tax Reform''.
        \item If incoherent/mixed: \textbf{"Unlabelable"}
    \end{itemize}
    \item \textbf{``description''}: A detailed explanation of the pattern, including specific nuances, emotional tones, or context \textbf{in English} (1-2 sentences).
    \item \textbf{``category''}: Choose one [Topic\_Specific, Entity\_Specific, Syntactic/Grammar, Pragmatic/Context, Polysemantic, Noise].
    \begin{itemize}
        \item If `Unlabelable', set this to \textbf{``Noise''} or \textbf{``Polysemantic''}.
    \end{itemize}
\end{enumerate}

Return ONLY the JSON string.
\end{tcolorbox}
\caption{The prompts provided to the LLM for automated feature annotation.}
\label{fig:prompt_details}
\end{figure}

\newpage

\section{Additional Case Studies on Latent Feature Steering}
\label{app:more_cases}

To further validate the precision of semantic steering, we present additional case studies across various domains. Each case highlights how deactivating a single monosemantic neuron (identified via our automated pipeline) redirects the retrieval mechanism to prioritize alternative semantic facets of the query.

In the following tables, the values in brackets represent the cosine similarity scores between the query (original or steered) and the retrieved sentences.

\subsection{Case Study 1: Neutralizing ``Jumping Activities'' (Neuron \#2039)}
The query \textit{``The record for the longest jump was broken during the Olympic trials''} initially retrieves sentences focused exclusively on jumping events. By clamping neuron \#2039, the focus shifts to the general concept of breaking records at Olympic trials across diverse sports.

\begin{table}[h]
\small
\centering
\begin{tabularx}{\textwidth}{l X X}
\toprule
\textbf{Rank} & \textbf{Original Retrieval (Top 5)} & \textbf{Steered Retrieval ($z_{2039}=0$)} \\ \midrule
1 & [0.8483] ...record) and triple jump (44 feet 0). & [0.8548] In 1900, Olympian Maxie Long set the first official world record in the 400 meters... \\
2 & [0.8449] Olympic skier Satre set a record jump of 112 feet (34 m) in 1937. & [0.8527] At the trials, she qualified for the finals... and broke the American record. \\
3 & [0.8448] The record was eventually broken... who jumped 75 meters (246 ft). & [0.8521] ...set a new Olympic record, beating Phelps' previous record of 51. \\
4 & [0.8445] ...Mike Powell, the world record holder in long jump. & [0.8495] ...Mike Powell, the world record holder in long jump. \\
5 & [0.8436] ...eventual gold-medal-winning (and Olympic record) jump of 6 feet 4. & [0.8452] ...breaking the Olympic record with a time of 21. / surpassed the Olympic A standard. \\ \bottomrule
\end{tabularx}
\caption{Retrieval results before and after deactivating the ``Jumping Activities'' feature.}
\end{table}

\subsection{Case Study 2: Neutralizing ``Bridge Infrastructure'' (Neuron \#6188)}
The query \textit{``The bridge construction was delayed due to an update in safety requirements''} is suppressed by deactivating the bridge-specific context, shifting the results toward general construction delays and safety compliance.

\begin{table}[h]
\small
\centering
\begin{tabularx}{\textwidth}{l X X}
\toprule
\textbf{Rank} & \textbf{Original Retrieval (Top 5)} & \textbf{Steered Retrieval ($z_{6188}=0$)} \\ \midrule
1 & [0.8647] The project was slated to have the 74-year-old bridge up to standards. & [0.8763] ...plans had to be revised to comply with new federal standards regarding steel pilings. \\
2 & [0.8646] To remedy what was becoming a major delay... bridge to cross the shipping channel. & [0.8745] Construction was delayed again a month later, with work to begin in February 2016. \\
3 & [0.8609] ...postponed until July 30, 2003, to improve safety on the highway. & [0.8710] Changes to the design and a lack of armor plating led to delays in building. \\
4 & [0.8602] Construction was delayed again... for completion in April 2017. & [0.8705] It has been updated to include regulations on ship construction and safety. \\
5 & [0.8601] ...plans had to be revised to comply with new federal standards. & [0.8691] But because the engineers needed to be recertified, the start was delayed again. \\ \bottomrule
\end{tabularx}
\caption{Retrieval results before and after deactivating the ``Bridge Infrastructure'' feature.}
\end{table}

\subsection{Case Study 3: Neutralizing ``Climate Change Awareness'' (Neuron \#9054)}
Suppression of the ``Climate Change Awareness'' feature in the query \textit{``The government announced a major policy shift regarding carbon tax credits''} shifts focus from environmental activism to general fiscal and administrative policy changes.

\begin{table}[h]
\small
\centering
\begin{tabularx}{\textwidth}{l X X}
\toprule
\textbf{Rank} & \textbf{Original Retrieval (Top 5)} & \textbf{Steered Retrieval ($z_{9054}=0$)} \\ \midrule
1 & [0.8524] The panel ultimately announced backing for a temporary carbon tax. & [0.8592] Its policy also represented a significant change from the idealism of previous governments. \\
2 & [0.8460] ...revised act called for participation in international carbon markets. & [0.8574] This effectively represented a complete rewrite of UK energy policy for the future. \\
3 & [0.8447] A carbon tax was introduced in 2012... but was scrapped in 2014. & [0.8545] ...the change was made because of privacy-related complaints. \\
4 & [0.8434] This effectively represented a complete rewrite of UK energy policy. & [0.8521] Sweden launched a new 10-year environmental tax shift. \\
5 & [0.8427] Abbott outlined his alternative climate change policy. & [0.8513] ...introduced a General Anti-Avoidance Rule to manage the risk of tax avoidance. \\ \bottomrule
\end{tabularx}
\caption{Retrieval results before and after deactivating the ``Climate Change Awareness'' feature.}
\end{table}

\end{document}